\documentclass[a4paper,10pt]{article}

\usepackage[a4paper,margin=0.9in]{geometry}

\usepackage{graphicx}      % For inserting images
\usepackage{amsmath}       % For mathematical symbols and equations
\usepackage{amssymb}       % For additional mathematical symbols
\usepackage{bm}            % For bold mathematical symbols
\usepackage{physics}       % For differential operators, bra-ket notation, etc.
\usepackage{mathtools}     % For additional math functionalities
\usepackage{hyperref}      % For hyperlinks
\usepackage[numbers,sort&compress]{natbib}
\usepackage{booktabs}      % For better table formatting
\usepackage{caption}       % For customizing captions
\usepackage{subcaption}    % For subfigures
\usepackage{xcolor}        % For colored text (if needed)
\usepackage{float}         % For better control of figure placements
\usepackage{enumitem}

\title{The central role of scientific goals in modeling complex\\ multiscale systems in nature}
\author{
Fabrizio Falasca \textsuperscript{1,*}\\[1ex]
\textsuperscript{1}Data Science Institute, University of Chicago, Chicago, USA \\
\textsuperscript{*}Corresponding author: \texttt{fabrifalasca@gmail.com}
}

\begin{document}
\maketitle
\begin{abstract}
The full complexity of multiscale turbulent systems in nature, such as Earth’s climate, exceeds what
any conceivable model can represent. Modeling such systems involves non-unique choices of which
variables and scales to resolve. Once such a coarse-grained representation is chosen, different models can be constructed to describe its evolution and account for the influence of unresolved processes. Which description is adequate depends on what we seek to predict or understand. We first illustrate this point in a controlled mathematical setting in which the resolved variables and scales are specified by construction. We fit a linear stochastic model to observations of the slow variables in a modified two-scale Lorenz--96 system. The choice of model structure sets limits on what it can predict even before fitting. For example, a linear model with additive noise cannot predict responses in ensemble variance to external forcing. Whether this limitation matters depends on the scientific question we aim to answer. The linear model represents stationary statistics of the non-linear system and it predicts accurately ensemble mean responses to forcings. The model therefore remains useful for predicting the mean response to forcing despite omitting the nonlinear interactions that mediate it in the original system. We then build on this specific example to examine the broader issue of how to construct and validate models of complex multiscale systems in nature. Such models are necessarily imperfect and limited. Thus, a model’s skill has no absolute meaning: it can be judged only relative to the observables of interest, which we select according to the study’s purpose. Model construction is therefore intertwined with the scientific question we seek to answer, and different questions about the same system often require us to build different models. The scientific purpose guides what each model must represent, which simplifications are acceptable, and how we judge its success.
\end{abstract}

%\noindent\textbf{Classification:} Applied Mathematics.\\
%\textbf{Keywords:} Coarse-graining | Modeling | High-dimensional chaosModeling | High-dimensional chaos
\tableofcontents

\section{Introduction}

The dynamics of high-dimensional turbulent systems differ substantially from those of familiar low-dimensional chaotic models. The latter often exhibit complicated invariant distributions and only a few unstable directions. In contrast, high-dimensional chaotic systems often feature numerous unstable directions, strong mixing, rapid decay of correlations, and multiscale dynamics \cite{MajdaIntroTurbulence}. Moreover, the stationary distributions of coarse-grained variables often show statistical regularities \cite{Colangeli2012}. Consequently, intuition drawn from low-dimensional chaos need not carry over directly to the effective, coarse-grained dynamics of turbulent systems. The study of climate dynamics illustrates this contrast. Earth's climate is a nonlinear, forced, and dissipative system with interacting processes spanning at least 15 orders of magnitude in space and time without a clear separation of scales \cite{GhilLucarini}. The challenge is compounded by having access to only partial observations of the system as well as by nonlinear interactions with unobserved degrees of freedom \cite{MajdaBranicki2012}. Modeling such systems inevitably involves non-unique choices of both coarse-graining representations, i.e. variables and scales, and modeling structures, including treatments of effective stochastic descriptions \cite{Hasselmann,LucariniChekroun,Dalmedico,FabriCoarseGraining}. In this work, we examine how specific scientific purposes guide these choices and determine which predictions must be tested to assess a model’s adequacy.\\

The widespread use of linear models in climate dynamics serves as a specific illustration of how the purpose of a model can guide the choice of its structure. Although Earth's climate is nonlinear, linear inverse models (LIMs) \cite{Penland89} have proved useful for describing selected statistics of low-frequency climate variability, see \cite{Penland95,PENLAND1996534,Penland2025,Capotondi,LIMOKane,LIMZanna,Villalobos,LIMManu} among many others. In part, the motivation comes from the fact that the stationary distributions of selected coarse-grained climate variables often exhibit a nearly Gaussian bulk \cite{MajdaStructuralStability}. Moreover, systematic stochastic mode reduction strategies have also shown that in some cases, such as barotropic quasi-geostrophic models, linear contributions and additive noise dominate the reduced dynamics \cite{FranzkeMajdaBarotropic}. The usefulness of linear and Gaussian approximations in large-scale climate dynamics extends beyond dynamical modeling to the problem of statistical and causal inference from data. Among the many successes: the Fluctuation-Dissipation Theorem (FDT), in its quasi-Gaussian approximation, has been used to predict the response of a state-of-the-art atmospheric circulation model to temperature perturbations with surprising skill \cite{GRITSUN}; Lütjens et al. (2025) \cite{Lütjens} showed that a linear regression-based emulator, akin to pattern scaling, can outperform deep learning emulators for the future climate projections they evaluated; information-theoretic quantities, including relative entropy and causation entropy, are often evaluated using Gaussian approximations \cite{majdaSIAM,nanChen,nanChenPhysicaD} with skillful results. Perhaps more surprisingly, Mascolo et al. (2025) \cite{Mascolo} showed that analytical Gaussian approximations of the committor function can yield skillful predictions of extreme events from coarse-grained data.\\

These successes do not establish that the effective coarse-grained dynamics are themselves linear: non-linear feedbacks and state-dependent noise remain important \cite{NormalForms}. They do, however, show that a model can accurately describe or predict selected observables without representing the full complexity of the underlying dynamics. This motivates two related lines of inquiry. First, we focus on the specific case of linear inverse models used to study large-scale climate modes. This class of models is often chosen despite its known limitations in representing the system's nonlinear dynamics. What motivates their adoption is their usefulness in describing and predicting relevant aspects of climate variability. We explore in a controlled setting the mechanisms that make linear modeling useful in a nonlinear multiscale system, and how relevant observables can be predicted with high accuracy. Second, we build on this example to develop a broader argument about modeling multiscale natural systems such as Earth's climate, whose full complexity exceeds what any conceivable model can represent \cite{MajdaBranicki2012}. In such systems, what we seek to predict or understand guides both the choice of variables and scales to resolve and the effective equations we write for their evolution.

\paragraph{An example of a useful imperfect model for selected observables.} We consider a modified Lorenz--96 system \cite{Lorenz96} with linear and quadratic couplings between the observed slow variables and the unobserved fast variables. We keep the number of resolved variables fixed and increase the number of unresolved fast variables. This setup parallels climate studies that model a few large-scale modes and average over a multitude of smaller, faster processes \cite{Penland95}. As the bath grows, the stationary distributions of the resolved variables become increasingly Gaussian. This statistical regularity emerges despite the strongly nonlinear dynamics of the full system, which in the largest case considered has approximately \(3{,}797\) positive Lyapunov exponents in a state space of dimension \(N=10{,}020\). Motivated by this regularity, we fit a linear stochastic model using only trajectories of the resolved variables \(\mathbf{x}\). The model accurately reproduces their stationary marginal distributions and lagged correlation functions. We then force the system with an additive forcing. Under such external forcing, the original system produces changes in both the ensemble mean and the ensemble variance. The model’s linear structure and additive noise rule out the observed variance response a priori \cite{Majda2010}. If the goal is to predict that change in variance, a more complex model structure is required \cite{NormalForms}. Despite this known limitation, the linear model is a justified choice when the goal is to predict the ensemble-mean response. The model is shown to capture this ensemble mean response accurately for both weak and strong perturbations.

\paragraph{\textit{Adequacy-for-purpose} in model construction and evaluation.} The Lorenz--96 experiment illustrates how the choice of model structure depends on the observable we seek to predict. An important simplification in idealized slow--fast models such as this one is that the slow variables \(\mathbf{x}\) are specified by construction. In a complex turbulent system such as Earth's climate, no clear separation of spatial and temporal scales singles out a unique set of variables to resolve \cite{GhilLucarini}. The scientific question therefore guides the choice of the coarse-grained representation itself, as well as the equations we write for its evolution \cite{FabriCoarseGraining}.\\
Building on \citet{Parker2020}, we argue that, for complex multiscale systems in nature, scientific purposes should guide which variables we resolve, how we model their evolution and account for the effects of unresolved degrees of freedom, and how we assess the model’s adequacy. What we seek to predict or understand helps determine which simplifications are acceptable, which predictions must be tested, and how accurate they need to be.\\ 
In data-driven climate modeling, the same dataset can support several models and emulators, each tailored to the observables and scales relevant to a particular scientific question, rather than expecting a single model to capture all aspects and complexity of the climate system.

%\section{Chaos in high-dimensions: an example, motivations and statistics}

%\section{High-dimensional chaos: a multiscale example and its statistics}

\section{Toy model: motivation, statistics, and high-dimensional chaos}

\subsection{Abstract formulation of turbulent dynamical systems} \label{sec:abstract}

To motivate our modified Lorenz--96 system and clarify its connection to climate dynamics, we begin with the general formulation of turbulent dynamical systems with quadratic nonlinearities \cite{Lorenz,MajdaIntroTurbulence}. Such formulation arises from projecting many fluid models onto orthogonal basis functions. It serves as a useful abstract formulation of turbulent dynamical systems. Denoting the state by $\mathbf{u}\in\mathbb{R}^N$, we write
\begin{equation}
\frac{\mathrm{d}\mathbf{u}}{\mathrm{d}t}
=
\mathbf{F}
+
\mathbf{A}\mathbf{u}
+
B(\mathbf{u},\mathbf{u}),
\label{eq:abstract_quadratic}
\end{equation}
where $\mathbf{F}$ is a forcing vector. The operator $\mathbf{A}$ represents the linear terms in the governing equations, which can include dispersion and dissipative processes. The quadratic term $B(\mathbf{u},\mathbf{u})$ arises from projecting nonlinear advection onto the same basis. These quadratic interactions are energy conserving, leading to the requirement
\begin{equation}
\langle \mathbf{u}, B(\mathbf{u},\mathbf{u})\rangle = 0
\label{eq:energy-conservation}
\end{equation}
where $\langle \cdot \rangle$ represents a suitable inner product \cite{MajdaEnergy}. Low-order models such as the Charney--DeVore model \cite{charneyDeVore} and the Lorenz--63, Lorenz--84 and Lorenz--96 models \cite{Lorenz,Lorenz84,Lorenz96} share this mathematical structure. The formulation also provides an abstract representation of atmospheric and oceanic dynamical dry cores \cite{MajdaBook}.\\
Continuum descriptions of turbulent dynamical systems are infinite-dimensional. Their discretized representations can contain very large numbers of degrees of freedom, i.e. $N \gg 1$. Coarse-grained modeling requires choosing which variables and scales to retain. We therefore partition the state $\mathbf{u}$ into resolved variables $\mathbf{x}$ at a chosen spatiotemporal resolution and unresolved variables $\mathbf{y}$, without necessarily assuming a clear separation of spatial or temporal scales. Following \citet{MTV1}, Eq. \eqref{eq:abstract_quadratic} is then rewritten by projecting onto the $\mathbf{x}$ and $\mathbf{y}$ variables as
\begin{equation}
\begin{aligned}
\frac{\mathrm{d}\mathbf{x}}{\mathrm{d}t} &= \mathbf{F}_1 + \mathbf{A}_{11} \mathbf{x} + \mathbf{A}_{12} \mathbf{y} + B_{11}^{1}(\mathbf{x},\mathbf{x}) + B_{12}^{1}(\mathbf{x},\mathbf{y}) + B_{22}^{1}(\mathbf{y},\mathbf{y}) \\
\frac{\mathrm{d}\mathbf{y}}{\mathrm{d}t} &= \mathbf{F}_2 + \mathbf{A}_{21} \mathbf{x} + \mathbf{A}_{22} \mathbf{y} + B_{11}^{2}(\mathbf{x},\mathbf{x}) + B_{12}^{2}(\mathbf{x},\mathbf{y}) + B_{22}^{2}(\mathbf{y},\mathbf{y}).
\end{aligned}
\label{eq:abstract_quadratic_slow_fast}
\end{equation}
A crucial issue with stochastic climate modeling is to (a) assume a scale separation between slow variables $\mathbf{x}$ and fast variables $\mathbf{y}$ and (b) define a new equation for the variability of $\mathbf{x}$ alone. The remaining fast variables $\mathbf{y}$ are replaced with new deterministic and stochastic terms \cite{MTV2,LucariniChekroun}. Crucially, while one could think that this issue can be avoided in new data-driven modeling and a deterministic neural network emulator could be built solely from observations of $\mathbf{x}$, the problem of taking into account the effects of the unobserved degrees of freedom keeps playing a role, as shown and discussed in \citet{FabriCoarseGraining} and later in Section \ref{sec:goals}.

\subsection{A modified Lorenz--96 model with linear and quadratic coupling} \label{sec:L96}
The two-scale Lorenz--96 system proposed by \citet{Lorenz96} is an idealized model of multiscale dynamics. It combines quadratic advection within the slow and fast subsystems with linear coupling between them. It has served as a toy model in the climate community to test new algorithms, see \cite{crommelinEric,BoffettaJAS,Arnold,M2LinesL96,giorgini2025kgmm} among many others. When fluid advection is projected onto a basis, quadratic interactions can couple resolved and unresolved modes, as in Eq.~\eqref{eq:abstract_quadratic_slow_fast}. We introduce quadratic cross-couplings into the two-scale Lorenz--96 system to represent such interactions in idealized form. The model uses a simplified coupling structure closely related to the nonlinear Lorenz--96 formulation of \citet{Abramov2013} and takes the following form:
\begin{equation}
\begin{aligned}
\frac{\mathrm{d}x_k}{\mathrm{d}t}
&=
x_{k-1}\left(x_{k+1}-x_{k-2}\right)
-x_k+F
+\frac{h_x}{J}(1+x_k)\sum_{j=1}^{J}y_{k,j},
\\
\frac{\mathrm{d}y_{k,j}}{\mathrm{d}t}
&=
\frac{1}{\epsilon}
\left[
y_{k,j+1}\left(y_{k,j-1}-y_{k,j+2}\right)
-y_{k,j}
+h_y\left(x_k+x_k^2\right)
\right].
\end{aligned}
\label{eq:l96_linear_quadratic_coupling}
\end{equation}
Here, $k=1,\ldots,K$ indexes the resolved variables, and $j=1,\ldots,J$ indexes the unresolved variables associated with each resolved variable. We choose $h_x=-1$ and $h_y=1$. With this choice of $h_x$ and $h_y$, the quadratic cross-couplings conserve the weighted energy \(E=\tfrac12\sum_k\bigl(x_k^2+\tfrac{\epsilon}{J}\sum_j y_{k,j}^2\bigr)\). The full quadratic operator satisfies Eq.~\eqref{eq:energy-conservation} in the inner product defined by \(E\). The parameter $\epsilon = 0.1$ controls the time scale separation. Eq. \eqref{eq:l96_linear_quadratic_coupling} uses periodic boundary conditions
\begin{equation}
x_{k+K}=x_k,
\qquad
y_{k+K,j}=y_{k,j},
\qquad
y_{k,j+J}=y_{k+1,j}.
\label{eq:l96_boundary_conditions}
\end{equation}
We denote the coupling tendency acting on the resolved variables by
\begin{equation}
B_k
=
\frac{h_x}{J}(1+x_k)\sum_{j=1}^{J}y_{k,j}.
\label{eq:l96_bath_forcing}
\end{equation}
The chosen parameters are
\begin{equation}
K=20,
\qquad F=16,
\qquad h_x=-1,
\qquad h_y=1,
\qquad \epsilon=0.1.
\end{equation}
We compare $J=1$, $3$, $10$, $500$ to examine how the stationary statistics change as the unresolved subsystem grows. The equations are invariant under cyclic shifts of the site index $k$. When analyzing the data, we exploit this translational symmetry by pooling stationary samples across the resolved sites to estimate probability distributions. The forcing value $F=16$ is associated with fully developed wave turbulence regime in the standard single-scale Lorenz--96 model \cite{MajdaQiControl}.

\subsection{Stationary statistics and high-dimensional chaos}

All four configurations are simulated using fourth-order Runge--Kutta integration with a timestep of $dt=0.005$. We retain $50{,}000$ model time units (MTU) after discarding an initial transient of $50$ MTU. Exponential fits $e^{-t/\tau}$ to the resolved autocorrelations, averaged across sites (because of translational invariance), give characteristic decay times of $\tau\approx0.108$ MTU for $J=1$ and $\tau\approx0.156$ MTU for $J=500$. The corresponding trajectories are then very long, spanning approximately $4.6\times10^5$ and $3.2\times10^5$ fitted decorrelation times, respectively.

\begin{figure}%[h!]
\centering
\includegraphics[width=1\linewidth]{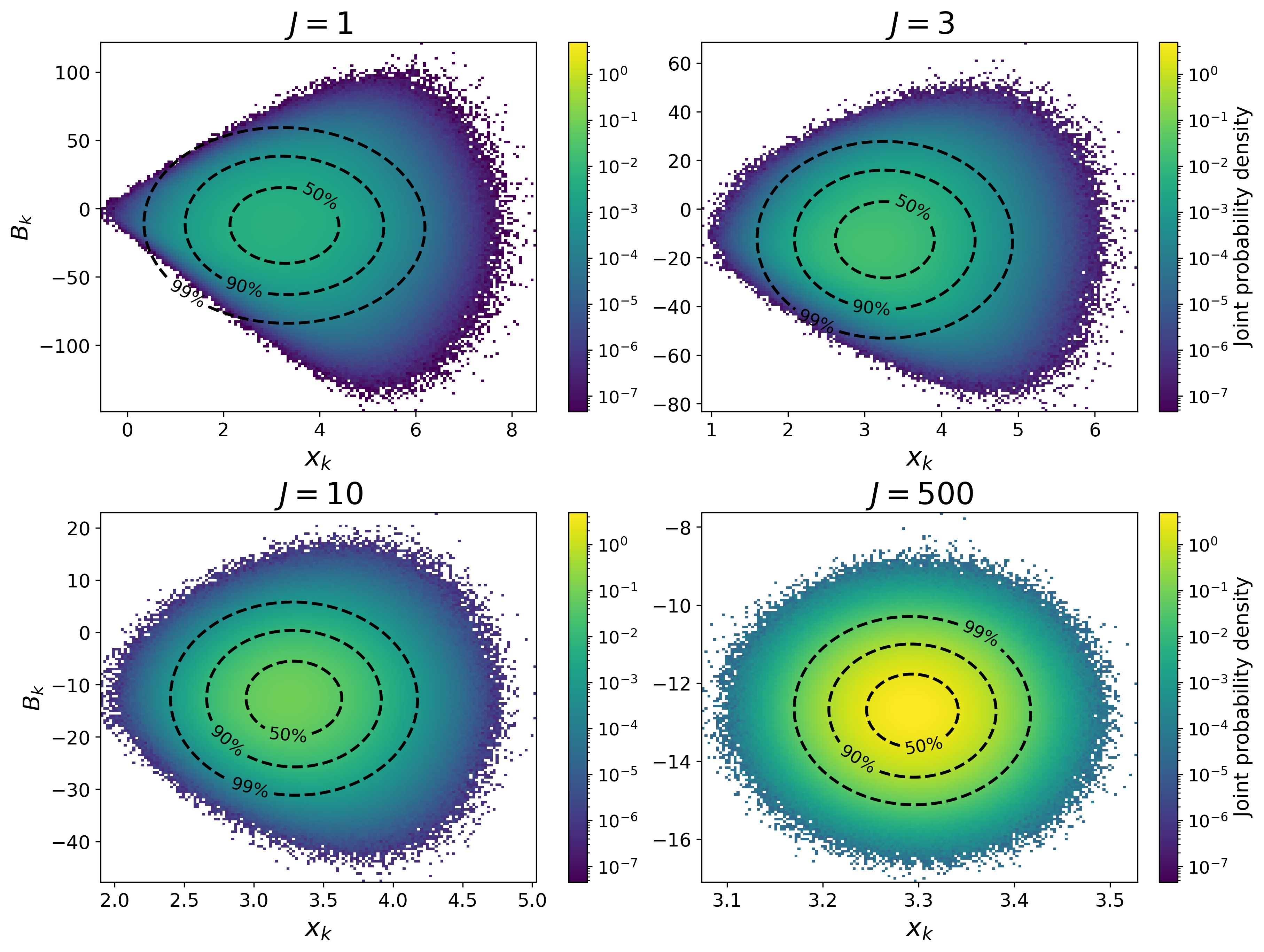}
\caption{Stationary joint probability densities of the resolved variable $x_k$ and coupling tendency $B_k$ for $J=1$ (top left), $J=3$ (top right), $J=10$ (bottom left), and $J=500$ (bottom right). The joint densities are estimated from trajectories pooled across the $K=20$ resolved sites (because of translational invariance). Dashed black lines show constant-density contours of a Gaussian with the same mean vector and covariance matrix as the data in each panel. The contours enclose the labeled probabilities ($50\%$, $90\%$, and $99\%$) under this Gaussian distribution. Increasing $J$ produces narrower joint distributions with progressively more Gaussian structure.}
\label{fig:Joint_PDFs}
\end{figure}

\paragraph{Gaussianization of joint statistics.} Figure~\ref{fig:Joint_PDFs} shows the stationary joint distributions of the resolved variable $x_k$ and the coupling tendency $B_k$. For $J=1$ and $J=3$, the distributions are strongly asymmetric and depart substantially from Gaussian distributions with matching means and covariances. Increasing $J$ progressively reduces this asymmetry: the central region becomes more elliptical at $J=10$, and the joint distribution is approximately Gaussian at $J=500$. Alongside this change in shape, the fluctuations of both $x_k$ and $B_k$ become markedly smaller.\\
When building reduced-order models of the slow variable $x_k$ alone the approximate joint Gaussianity motivates the use of simplified stochastic closures as $J$ increases. In fact, under a joint Gaussian approximation, the conditional mean of $B_k$ depends linearly on $x_k$, and its conditional variance is independent of $x_k$:
\[
\mathbb{E}[B_k\mid x_k]\underset{J\gg1}{\approx} a+b x_k,
\qquad
\operatorname{Var}(B_k\mid x_k)\underset{J\gg1}{\approx}\sigma^2.
\]
After accounting for the conditional mean, the remaining fluctuations can be approximated by an additive Gaussian term at a given instant. $\sigma$ is a constant. These equal-time statistics do not, however, determine the temporal dependence of the residual forcing, which may require a model with memory.\\
As the coupling $B_k$ involves an average over many fast variables, the observed Gaussianization suggests a central-limit-type mechanism. However, in our experiments, increasing $J$ in the original linearly coupled Lorenz--96 system does not produce comparable Gaussianization (not shown). This suggests that the statistical simplification depends on the combined effects of a large unresolved bath and nonlinear, energy-conserving coupling between resolved and unresolved variables. Including these quadratic couplings brings the model closer to the interaction structure of turbulent dynamics represented by the abstract formulation in Section~\ref{sec:abstract}.\\ 
The cumulative effect of many unresolved processes may likewise contribute to the nearly Gaussian statistics of some large-scale climate variables. In a related study of a barotropic quasi-geostrophic model, \citet{MajdaMetastable} showed that strongly nonlinear interactions can coexist with nearly Gaussian stationary distributions of large-scale modes.

\paragraph{High-dimensional chaos.}
The nearly Gaussian resolved statistics at $J=500$ emerge from a nonlinear system with thousands of unstable directions. Using the Benettin algorithm with periodic orthonormalization of tangent vectors through QR decomposition \cite{Benettin}, we estimate approximately $3{,}797$ positive Lyapunov exponents in a state space of dimension $N=K(1+J)=10{,}020$. The estimated unstable subspace therefore accounts for approximately $38\%$ of the full dimension. This coexistence of statistical simplicity and high-dimensional chaos motivates our focus on $J=500$. It also provides an idealized setting for understanding the effectiveness of linear inverse models in climate applications \cite{PENLAND1996534}: selected statistics of coarse-grained variables may admit useful linear stochastic descriptions even when the underlying dynamics are nonlinear and strongly chaotic. In the next section, we examine how well a data-driven linear stochastic model, fitted solely to observations of the $20$ resolved variables, reproduces their variability and responses to external forcing.

\section{An imperfect but useful data-driven reduced-order model} \label{sec:LLM_experiment}

Using the stationary trajectory of the modified Lorenz--96 system with $K=20$ and $J=500$, we construct a reduced-order model from observations of the resolved variables $\mathbf{x}$ alone. We first describe its formulation and estimation, then evaluate its stationary statistics and responses to weak and strong external forcings.

\subsection{Linear model with memory (LMM)}

Following the multilevel modeling approach of \citet{KONDRASHOV201533}, we consider a linear stochastic model with two memory layers:
\begin{equation}
\begin{split}
\begin{aligned}
\mathrm{d}\mathbf{x} &= \left(\mathbf{c}_0+\mathbf{L}\mathbf{x}+\mathbf{r}\right)\mathrm{d}t \\
\mathrm{d}\mathbf{r} &= \left(\mathbf{c}_1+\mathbf{A}\mathbf{x} +\mathbf{B}\mathbf{r}+\mathbf{s}\right)\mathrm{d}t \\
\mathrm{d}\mathbf{s} &= \left(\mathbf{c}_2+\mathbf{C}\mathbf{x} +\mathbf{D}\mathbf{r}+\mathbf{E}\mathbf{s}\right)\mathrm{d}t +\boldsymbol{\Sigma}\mathrm{d}\mathbf{W}.
\end{aligned}
\end{split}
\label{eq:LMM}
\end{equation}
Here, $\mathbf{r}$ and $\mathbf{s}$ are auxiliary variables representing memory effects, $\mathbf{W}$ is a standard Wiener process, and all matrices and constant vectors are estimated from observations of $\mathbf{x} = (x_1, ..., x_{20})$ alone.\\
We fit the model using data sampled every $\Delta t=0.05$ MTU. For numerical conditioning, we standardized data by subtracting their empirical means and dividing by their standard deviations; the fitted coefficients are then converted back to the original units.\\
The fitting procedure is based on simple ordinary least-squares regressions. First, we fit the resolved tendency and define its residual as $\mathbf{r}$. We then fit the tendency of $\mathbf{r}$ using $(\mathbf{x},\mathbf{r})$, obtaining the residual $\mathbf{s}$. Finally, we fit the tendency of $\mathbf{s}$ using $(\mathbf{x},\mathbf{r},\mathbf{s})$. The model is then closed as a Wiener process $\boldsymbol{\Sigma}\mathrm{d}\mathbf{W}$: we consider the residuals of the last layer and compute the noise matrix $\boldsymbol{\Sigma}$ via Cholesky decomposition of the residual covariance matrix. Tendencies are estimated using forward differences, and the fitted model is simulated using Euler--Maruyama at the same interval $\Delta t = 0.05$. Notably, the model's approximation comes from both its linear formulation and the Euler--Maruyama integration at the finite time step $\Delta t$.\\
We estimate all coefficients using the full retained stationary trajectory (without training/validation splitting). The fitted model is then validated in two steps: (a) through long simulations, comparing its stationary PDFs, autocorrelations, and cross-correlations with those of the original system; (b) through its skill in predicting responses to weak and large external forcings. The second test is quite demanding as no perturbed trajectories are used in fitting.

\subsection{Stationary statistics}

\begin{figure}%[h!]
\centering
\includegraphics[width=1\linewidth]{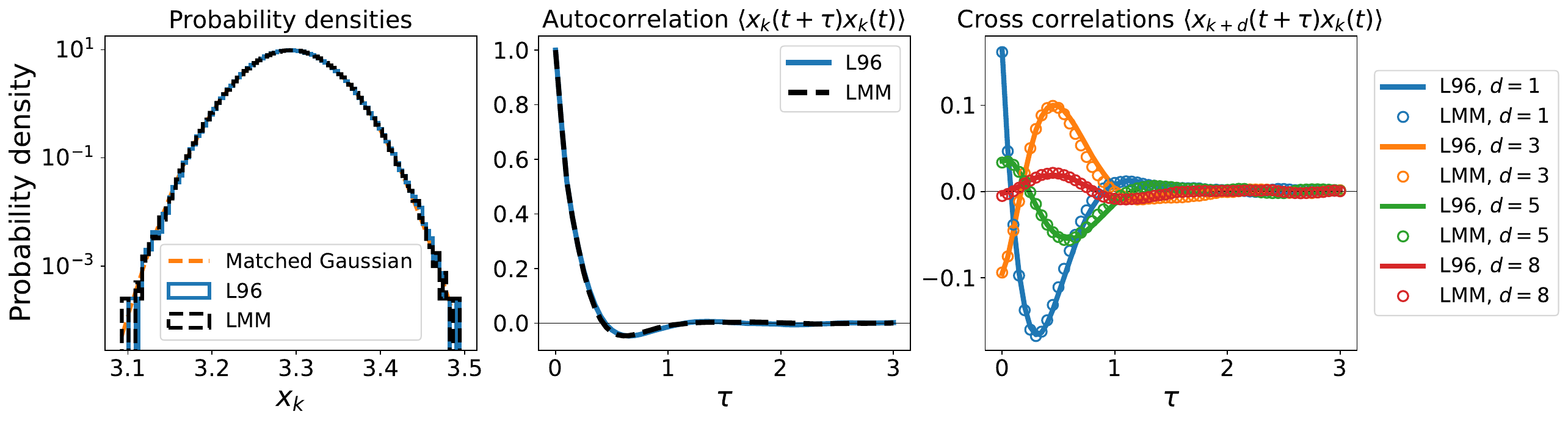}
\caption{Stationary statistics of the modified Lorenz--96 system (L96) and the linear model with memory (LMM) for $K=20$ and $J=500$. Left: marginal probability densities of $x_k$ for L96 and the LMM, compared with a Gaussian matching the L96 mean and variance. Center: lagged autocorrelation functions. Right: lagged cross-correlation functions between $x_{k+d}(t+\tau)$ and $x_k(t)$ for separations $d=1,3,5,8$. Lags $\tau$ are expressed in model time units.}
\label{fig:PDF_correlations}
\end{figure}

Using the fitted LMM in Eq.~\eqref{eq:LMM}, we generate a stationary trajectory of $50{,}000$ model time units (MTU), matching the duration of the full-model simulation. The LMM describes the $K=20$ resolved variables $\mathbf{x}$ alone. We compare the marginal distributions of $x_k$, their lagged autocorrelation functions, and the cross-correlation functions between $x_{k+d}(t+\tau)$ and $x_k(t)$ for four different $d$. All correlations are computed from centered, normalized variables. Given the translational invariance of the original system, all $x_k$ share the same statistics. 
Figure~\ref{fig:PDF_correlations} shows that the LMM accurately reproduces the marginal distributions. It also captures the rapid decay and weak negative lobe of the autocorrelation function. The cross-correlation functions also agree closely across all spatial separations; four representative examples are shown.

\subsection{Perturbed statistics}

\subsubsection{Linear response to an impulse perturbation} \label{sec:response_operator}

\begin{figure}%[h!]
\centering
\includegraphics[width=1\linewidth]{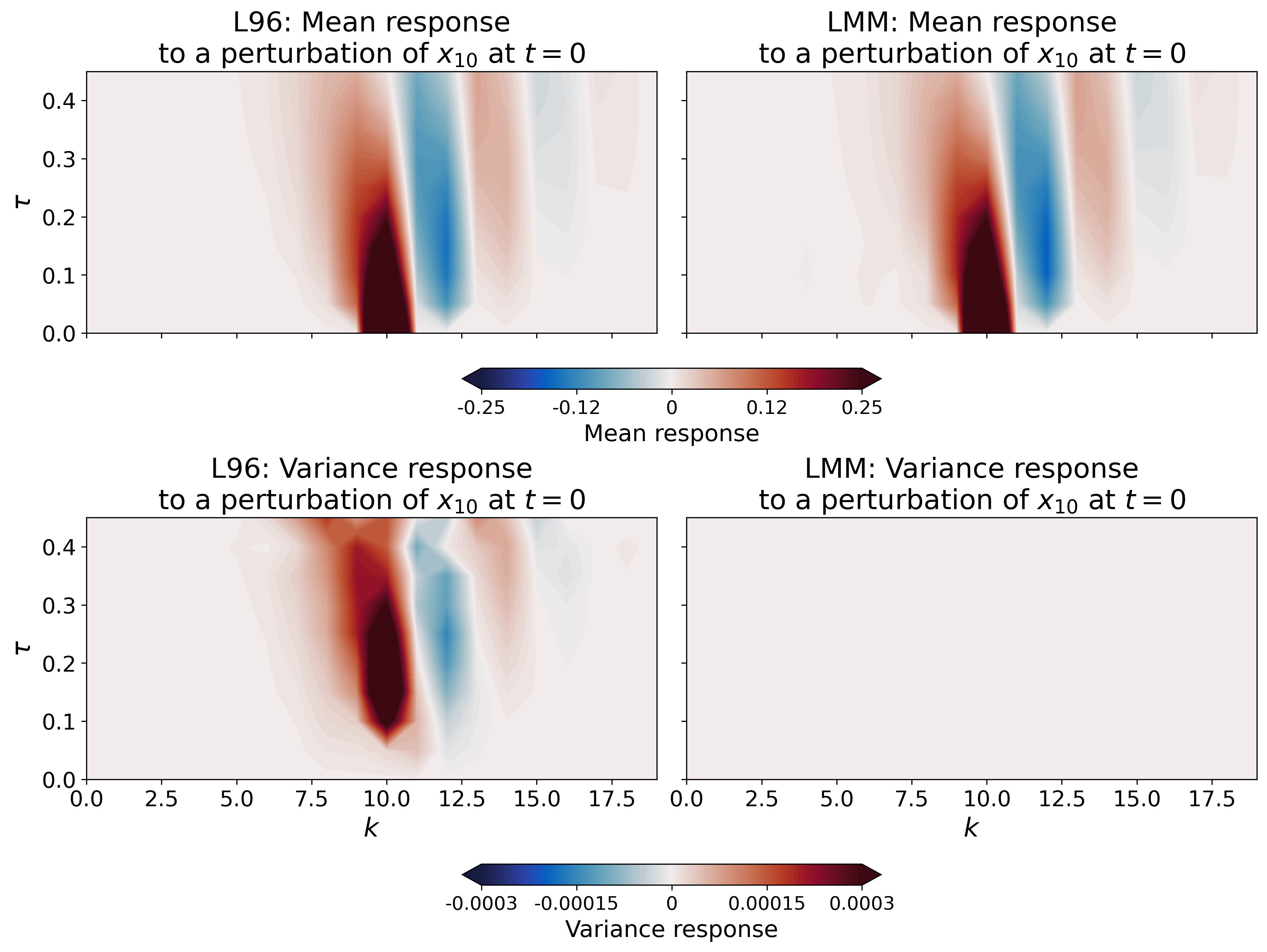}
\caption{Modified Lorenz-96 system: Linear response regime. Response operator $R_{m,l}(\tau)$ for the ensemble mean responses (top row) and variance responses (bottom row) to a small impulse perturbation of $x_{l = 10}$ at $t=0$. Left column: modified Lorenz--96 system with $J=500$. Right column: fitted data-driven LMM. The LMM closely reproduces the mean response but predicts zero variance response, as expected.}
\label{fig:Impulse_Response}
\end{figure}

We apply a small impulse perturbation $\delta x_l(0)$ at site $l$ and at time $t=0$. Starting from a large ensemble of initial conditions sampled from the stationary regime, we evolve pairs of control and perturbed trajectories and measure the resulting changes in ensemble mean and variance. More formally, we focus on the responses of two observables: $\mathcal{A}(x_m(\tau))=x_m(\tau)$ denoting the mean; $\mathcal{A}(x_m(\tau))=(x_m(\tau)-\mu_m(\tau))^2$ denoting the variance, where $\bm{\mu}(\tau)$ represents the time-dependent mean. The time-dependent response operator of interest here is defined as:
\[
R_{m,l}(\tau) = \frac{\delta \langle \mathcal{A}(x_m(\tau)) \rangle}{\delta x_l(0)}.
\]
where $\delta \langle \mathcal{A}(x_m(\tau)) \rangle$ indicates the response (in terms of ensemble averages) of the observable to the impulse perturbation $\delta x_l(0)$. In this work we apply an impulse perturbation on site $x_l = x_{10}$ and analyze the change in mean and variance caused by this perturbation. In practice, very large ensembles are needed to compute these responses (order $10^8$ in our case for reliable computation of ensemble variances) and we report the computational steps in Appendix \ref{app:app_Response_Operator}.\\
Results are reported in Figure \ref{fig:Impulse_Response}. The mean response develops alternating positive and negative lobes, whose spatial structure and temporal evolution are closely reproduced by the LMM. Validation based solely on this agreement could give the misleading impression that the model has recovered the underlying effective dynamics. The variance response reveals the model's limitation: the modified Lorenz--96 system exhibits a nonzero response, indicating that the perturbed distribution changes beyond a shift in its mean. The LMM cannot reproduce this behavior: a linear model with additive noise cannot reproduce responses in variance to forcing. This structural limitation is known a priori \cite{Majda2010}. Thus, the model accurately captures the mean response while failing to capture the full distributional response to the same perturbation.\\

\subsubsection{Response to a large step function forcing}

\begin{figure}%[h!]
\centering
\includegraphics[width=1\linewidth]{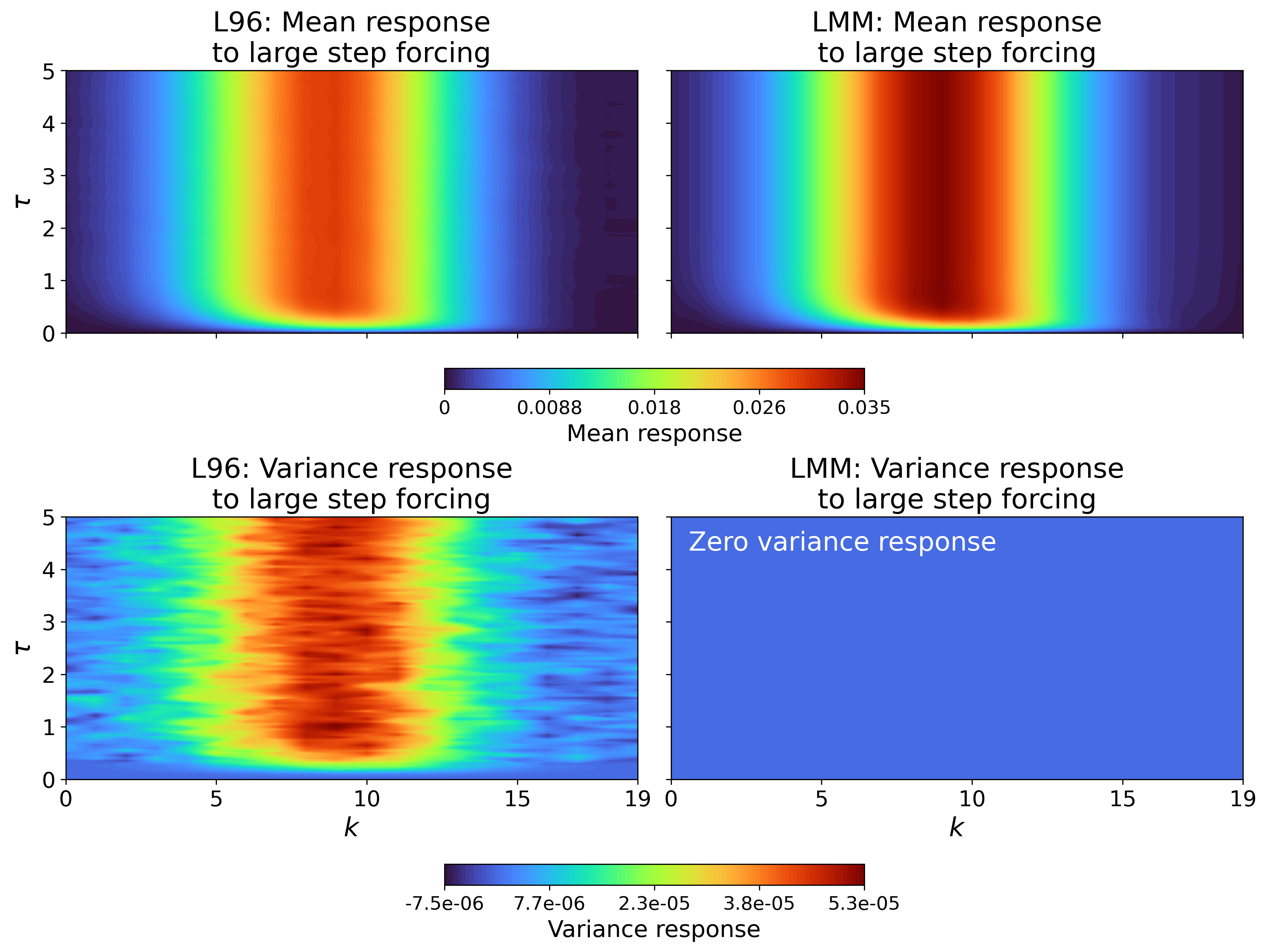}
\caption{Modified Lorenz-96 system: Response to a large sustained forcing. Response of the numerical and LMM models to a large Gaussian-shaped forcing Eq. \eqref{eq:gaussian_forcing} applied and held constant for $t \geq 0$. Top row: ensemble-mean response. Bottom row: ensemble-variance response.}
\label{fig:Step_Forcing}
\end{figure}

We now examine the response to a large, sustained forcing applied to the slow variables. We add a ``Gaussian bump'' forcing $G_k(t)$ to the right-hand side of the $x_k$ equations in Eq.~\eqref{eq:l96_linear_quadratic_coupling} and to the first level of the LMM in Eq.~\eqref{eq:LMM}. The forcing is switched on at $t=0$ and held constant thereafter:

\begin{equation}
G_k(t)
=
A \exp\left[-\frac{d(k,k_0)^2}{2\ell^2}\right] ~ \text{for} ~ t \geq 0,
\qquad
d(k,k_0)
=
\min\left(|k-k_0|,\,K-|k-k_0|\right).
\label{eq:gaussian_forcing}
\end{equation}
Here, $d(k,k_0)$ accounts for the periodic boundary conditions. We set $k_0=10$, $\ell=3$, and $A=5\sigma_{\mathrm{ref}}$; $\sigma_{\mathrm{ref}}$ being the standard deviation of $x_{k}$ in the unperturbed system. This defines a severe forcing amplitude, of the order of 5 times the standard deviation of the variable $x_k$ in the Lorenz--96 equations. We integrate forced and unforced ensembles and measure the differences in their ensemble means and variances. We use $10^6$ ensemble members for both the LMM and the numerical model.\\ 
Results are shown in Figure~\ref{fig:Step_Forcing}. The forcing produces a change in both the ensemble mean and variance of the modified Lorenz--96 system, clearly identifying changes in the position and shape of the attractor in state space. The LMM captures the spatial structure and temporal adjustment of the mean response quite well, although it slightly overestimates its amplitude. However, its ensemble-variance response is identically zero, as expected from theoretical arguments \cite{Majda2010}.\\

\paragraph{Does the linear approximation remain useful at \(J=1\)?} The nearly Gaussian stationary marginals of the resolved variables at \(J=500\) motivated our choice of a linear stochastic model. Even when these marginals are non-Gaussian, a linear model may still capture lagged correlations and ensemble-mean responses to forcing. Appendix~\ref{app:experiment_J_1} shows that at \(J=1\), the fitted LMM misses the non-Gaussian shape of the stationary marginals but still captures the lagged correlations and the main features of the ensemble-mean response to both the impulse and the sustained forcing.

\section{Scientific goals in model construction and evaluation} \label{sec:goals}

\subsection{Validating imperfect but useful descriptions of complex systems}

In the preceding experiment, we found that the stationary marginal distributions of the 20 slow variables of a modified Lorenz--96 model became more Gaussian as we increased the number of unobserved fast variables. This motivated our choice of a linear model with memory (LMM) for the case \(J=500\). The LMM reproduces the stationary marginal distributions and lagged correlations of the slow variables. It also captures the spatial pattern and temporal evolution of their ensemble-mean response to additive forcing. Yet the same forcing produces a significant change in ensemble variance in the modified Lorenz--96 system. We knew before fitting that the LMM’s linear structure and additive noise rule out such a variance response to forcing \cite{Majda2010}. Thus, even for the same forcing, whether the LMM is adequate depends on which observable we seek to predict.

\paragraph{Capturing the variance response in the modified Lorenz--96 system.} Nonlinear corrections to the deterministic drift and state-dependent noise would allow us to capture a response in variance to external forcing \cite{MTV2,NormalForms}. However, predicting the magnitude and time evolution of that response is difficult. For this task, the model must account for nonlinear interactions between observed and unobserved degrees of freedom: a model fitted to unforced observations of the slow variables alone must capture how the unobserved fast variables influence ensemble spread under forcing \cite{MajdaBranicki2012}. In the modified Lorenz--96 system, formulating effective stochastic equations for the slow variables is a tractable, well-defined problem because there are only two scales and the slow variables are known exactly. For such idealized systems, an extensive literature describes analytical and data-driven approaches to modeling the effective dynamics and predicting responses to forcing \cite{Abramov2013,SapsisMajdaPNAS,SapsisPoF,GiorginiScore1,GiorginiScore2,stableSI}. This is not the case in  multiscale systems in nature.

\paragraph{Choosing and validating an effective description of complex multiscale systems in nature.} In complex turbulent systems such as the Earth's climate, processes interact across a broad range of spatial and temporal scales with no clear scale separation \cite{GhilLucarini}. Deciding which variables and scales to resolve is therefore itself a modeling choice. Such choices are non-unique and mainly guided by what we seek to predict or understand \cite{FabriCoarseGraining}. Even after this choice is made, models with different parametrizations of the unobserved processes can agree closely on stationary statistics while predicting different responses to the same external forcing \cite{majdaSIAM}. Each possible model choice is an imperfect description of the full system. What we can establish is only how accurately it predicts selected observables under specified conditions. For example, a data-driven weather model can produce skillful forecasts while underestimating the growth of small initial perturbations \cite{hassanzadeh2026missing}. Similarly, a nonlinear deterministic climate emulator, inevitably trained on coarse-grained observations, could predict the ensemble-mean response to forcing perfectly and still miss the variance response. The scientific question determines which of these limitations matter for the model’s intended use and which predictions must be tested to establish its adequacy.\\

In what follows, we first examine examples of how different scientific goals can favor different strategies for climate model development. We then draw on the LMM experiment to discuss broader implications for model construction and validation. For clarity: developing new modeling methods is a valuable scientific goal in its own right. Given a set of existing methods, we ask how the scientific question should guide their selection and use in building a model.

\subsection{Different scientific goals can favor different strategies. Examples from climate modeling}

Increasing resolution and representing more processes explicitly have long been important strategies in the development of general circulation models (GCMs) \cite{Pacchetti}. These developments can advance process understanding, but do not by themselves establish greater reliability in climate projections. For example, models became more complex between phases 3 and 5 of the Coupled Model Intercomparison Project (CMIP3 and CMIP5), yet there was little evidence that large-scale climate feedbacks were better constrained \cite{Knutti2013}. Similarly, despite improvements in the representation of individual physical processes, some CMIP6 models overestimated the observed warming trend over 1981--2014, suggesting that they may also overestimate future warming under a given emissions scenario \cite{Tokarska}.\\
Increasing resolution or the system's complexity can also limit the number of ensemble members that can be run \cite{Pacchetti}. Climate projections are inherently statistical, and reliable estimates of forced changes in means and variances require sufficiently large ensembles. Estimating changes in variance can be particularly demanding, and even at conventional resolutions, an ensemble may be too small to distinguish a forced change in variance from sampling fluctuations. Increasing resolution can therefore advance the goal of understanding particular physical processes while working against the goal of producing reliable probabilistic projections \cite{Pacchetti}. A useful alternative is to retain the existing resolution and improve the parametrization of the unresolved processes that matter most for a given projection \cite{Schneider2024,Zanna2025}. A modeling program can pursue both goals through complementary configurations, each evaluated against the question it was designed to answer.\\ 

Comprehensive machine-learning (ML) emulators are being developed to complement numerical climate simulations \citep[e.g.,][]{Camulator,Samudra,Samudra2,Watt-Meyer2024,Lucie}. These models also require explicit criteria for success. The aim of reproducing as many aspects of a target system as possible leaves open which predictions must be reliable, under which conditions, and to what accuracy. Specifying the intended use helps identify both what must be tested and what to emphasize during training. The same dataset can thus support many data-driven models, from emulators that predict successive weather fields to reduced-order models that evolve large-scale modes directly. For example, if the goal is ENSO prediction, the model can be built around the variables and scales relevant to ENSO variability, without having to predict the sequence of weather fields from which it arises \cite{KondraSeasonality}.\\ 
Dynamical emulators, or their successors, could eventually serve as numerical laboratories for investigating specific causal relationships and advancing our mechanistic undertanding of the system. Physical constraints can guide their construction \cite{WillG}, and response theory can provide benchmarks for validating their responses \cite{FabriCoarseGraining,FabConstraints}. Part of the appeal is that, for many causal and sensitivity questions, it is enough to apply a small perturbation and trace its effects near the attractor sampled during training. However, even within this regime the choice of emulator depends on the causal question at hand. For example, if we want to trace how a localized perturbation excites atmospheric waves and how those waves carry its influence downstream, we might use an emulator that evolves gridded weather fields and resolves wave propagation. If we want to estimate how perturbing a large-scale mode of climate variability changes another mode over time, a reduced-order model that evolves those modes directly may suffice. If instead the interest is in projections involving larger departures from the reference climate, statistical emulators trained on output from comprehensive climate models could offer an efficient way to estimate how selected observables change across forcing scenarios \cite{Lütjens,AndreGenerative}. Forecasting ENSO, tracing the propagation of a perturbation, and projecting changes across forcing scenarios can therefore call for different emulators, each designed and tested for its intended use.

\subsection{Some implications for process-based and data-driven modeling}

In our experiment in Section \ref{sec:LLM_experiment}, changing the observable from the ensemble mean to the variance changes our assessment of the very same model. This gives concrete meaning to George Box's observation that ``all models are wrong, but some are useful''. What, specifically, do we want a model to help us predict or understand? Asking this question at the outset allows the answer to guide both model construction and validation.\\

Building on \citet{Parker2020,FabriCoarseGraining,Thompson}, we discuss four implications for process-based and data-driven modeling of multiscale systems, with particular attention to Earth's climate dynamics. We use the LMM experiment to illustrate their practical consequences.

\paragraph{Scientific purposes guide model construction.}
Scientific goals guide the choice of resolved variables and scales \cite{FabriCoarseGraining}. In a data-driven model, this choice determines what dynamics the model is asked to learn and which processes require a closure. The required treatment of unresolved processes also depends on the observables of interest. A deterministic Markovian emulator may reproduce ensemble-mean responses while missing effects of unresolved variability that matter for variance responses \cite{MajdaBranicki2012}. Success in predicting the mean response can therefore conceal the need for stochastic parametrizations or other closures \cite{FabriCoarseGraining}. Physical and mathematical knowledge can guide this choice before fitting. In our LMM, linear dynamics and additive noise imply that deterministic additive forcing leaves the ensemble variance unchanged. If the scientific goal includes predicting the observed change in variance, a different model structure is therefore required.

\paragraph{Scientific purposes guide model validation.}
A ``mirror'' view judges model quality by how comprehensively and accurately a model represents its target system \cite{Parker2024}. Yet a model can perform well across many diagnostics while failing on the predictions that motivated its construction. Scientific goals must therefore determine what evidence is relevant to model validation \cite{Parker2020}. For example, agreement with stationary statistics is insufficient to establish the accuracy of a model's response to forcing \cite{majdaSIAM}. A process-based climate model may reproduce the observed statistics via parameter tuning while still misrepresenting how those statistics change under forcing. Similarly, a data-driven climate model can reproduce the statistics of its training data while giving inaccurate responses to perturbations \cite{FabriCoarseGraining}. For both classes of model, linear response theory can provide benchmarks for responses to small external perturbations. These benchmarks can be estimated from unperturbed observations of the target system, without knowing its governing equations or performing perturbation experiments \cite{FabConstraints,Boger,wettlauferChaos}. Even within the linear response regime, infinitely many forcing patterns could be tested. The scientific goal determines which perturbations to consider, which observables to examine, and how accurately their responses must be reproduced.

\paragraph{Different purposes can require different models for the same physical system.} Predicting large-scale anomalies and explaining a physical feedback can require different variables, resolutions, and closures. These differences motivate complementary models and the study of their relationships within model hierarchies \cite{HeldGap,Jeevanjee,Shaw2026}. For example, \citet{GhilRobertson2002} discuss how coarse-grained descriptions of low-frequency atmospheric variability in terms of persistent regimes and intraseasonal oscillations complement efforts of detailed numerical models. Manabe’s work on radiative-convective equilibrium illustrates how a deliberately simplified model can yield predictions and physical insight into climate sensitivity \cite{Manabe}. Data-driven reduced-order models have likewise advanced understanding of climate dynamics by focusing on selected variables, scales, and processes \cite{KRAVTSOV,KondraSeasonality,STROUNINE2010145,ChenENSO,KITSIOS,KITSIOSOcean,Wettlaufer1,LudoWettlaufer,FabConstraints}. The same physical system can thus admit several imperfect but useful descriptions, each suited to a particular scientific question. The question guides our choice of variables, scales, and processes, and those choices in turn shape the effective dynamics the model can represent.

\paragraph{What can a model's predictive success tell us about the underlying dynamics?}
Establishing which physical mechanisms a data-driven model has captured requires evidence beyond its success in predicting selected observables \cite{LennyThorpe}. In Section~\ref{sec:response_operator}, we compare the evolution of the mean-response operator of the modified Lorenz--96 system with the one predicted by the LMM. The response operator quantifies how a small \textit{external} perturbation to one variable changes the ensemble mean of another variable at later times. These responses quantify time-dependent causal links in the interventional sense, as argued by \citet{Aurell,Lucarini2018,Baldovin}. The LMM is useful for estimating these causal effects on the ensemble mean, even though its equations omit the nonlinear interactions that generate them in the original system. Predicting the effect of an intervention therefore leaves open which physical mechanisms mediate it.

\section{Conclusions}

Modeling turbulent natural systems such as Earth's climate inevitably leads us to work with partial, coarse-grained observations of high-dimensional dynamics spanning many interacting scales \cite{MajdaBranicki2012,GhilLucarini}. A central challenge is therefore to describe selected variables while accounting for their interactions with processes we do not observe or resolve. These interactions can require deterministic corrections, stochastic parametrizations, and memory in the effective description. Neither the choice of variables and scales nor the treatment of unresolved processes is unique: both must be guided by the scientific questions and observables of interest \cite{FabriCoarseGraining}. This motivates hierarchies of complementary models \cite{HeldGap}, whose construction and evaluation depend on the purposes they are intended to serve.\\

In the first part of the paper, we introduced a modified two-scale Lorenz--96 system with linear and quadratic coupling between resolved and unresolved variables. Increasing the number of unresolved variables produces increasingly Gaussian resolved statistics. At $J=500$, this statistical simplicity coexists with thousands of positive Lyapunov exponents. This statistical regularity motivated us to test a linear stochastic model for a fixed set of resolved variables, drawing on the use of linear inverse models for large-scale climate anomalies. The linear model was fitted solely to coarse-grained observations: the model reproduces selected stationary statistics and captures the ensemble-mean responses to weak and strong perturbations. Its linear structure and additive noise rule out the observed variance responses a priori. Despite this known limitation, the linear model is a justified choice when the goal is to predict the ensemble-mean response. The same model is therefore useful for some observables and inadequate for others. Developing a closure that predicts the variance response is challenging but tractable in this two-scale system with known slow variables. The task is considerably harder in multiscale turbulent systems in nature, where the choice of variables and scales to resolve is itself part of the modeling problem.\\

We used this example to examine model construction, evaluation, and interpretation from an \textit{adequacy-for-purpose} perspective \cite{Parker2020}. The complexity of a physical system does not, by itself, determine the model complexity required for a particular task. This distinction is particularly important when interpreting the success of data-driven models: reproducing selected statistics or even some responses to external interventions does not establish that a model has learned the underlying physical mechanisms. Nor does it establish reliability for other observables or forcing conditions. For complex, partially observed systems such as Earth's climate, these considerations support complementary modeling approaches addressing different scales and processes of interest \cite{Thompson,Pacchetti}. Comprehensive numerical models remain valuable tools for investigating interacting physical processes and their responses to forcing. However, neither increasing the physical detail and complexity of numerical models nor building general-purpose data-driven emulators establishes greater adequacy for a given scientific purpose. For example, as we argued, increasing resolution may improve our understanding of a physical process while making it impractical to run the large ensembles needed to estimate forced changes in variability.\\

How we build a model is inseparable from the scientific question we want to answer. Different questions can therefore justify building several imperfect models of the same physical system. In practice, a central modeling task is deciding what not to resolve explicitly. Predicting a hurricane’s landfall, for example, does not require tracking individual molecules. The continuum equations used for this purpose are themselves an effective description in terms of fluid fields that average over molecular motion. Likewise, selected statistics of large-scale climate modes can be captured by modeling those modes directly, without explicitly resolving individual weather events. The scientific purpose guides what each effective description must represent, what it can simplify, and how its success should be assessed.

\paragraph*{Acknowledgments.}
This work was supported in part by NSF grant AGS-2531264 and by the AI for Climate Initiative (AICE) at the University of Chicago's Data Science Institute. Computational resources and support were in part provided by the NYU IT High Performance Computing facilities, services, and staff. Research was partially completed while the author was visiting the Institute for Mathematical Sciences, National University of Singapore in 2026.

\appendix

\section{Numerical estimation of model response operators} \label{app:app_Response_Operator}

To estimate the response operator we follow the steps proposed in \cite{FabriCoarseGraining}. Given a model $\dot{\mathbf{x}} = \mathbf{f}(\mathbf{x}_t) + \mathbf{\Sigma}\bm{\xi}(t)$, we build the impulse response operator as follows:

\begin{enumerate}
    \item We simulate a very long trajectory from a random initial condition and remove an initial transient. We then sample $N_e$ random points from the simulation to define an ensemble of $N_e$ initial conditions on the model’s attractor. $N_e$ should be very large (i.e. $N_e \gg 1$) in order to sample the whole attractor and approximate averages over the invariant distribution.
    \item For each one of the $N_e$ initial conditions, we impose an impulse perturbation $\Delta_l$ to the degree of freedom $x_l$ at time $t = 0$. The amplitude of the perturbation  should be theoretically infinitesimally small, ensuring linearity of the response even in nonlinear systems. In practice we do as follows: we consider the long time series of $x_l(t)$ from the long control integration above and define $\Delta_l = 10^{-1} \sigma_l$, where $\sigma_l$ is the standard deviation of time series $x_l(t)$. 
    \item Therefore, for a given initial condition $\mathbf{x}(t=0)$, we simulate two trajectories: a control trajectory without perturbation and a perturbed one, where an impulse perturbation $\Delta_l$ has been imposed on the $l$-th degree of freedom at time $t = 0$. This procedure is repeated in parallel for all initial conditions, resulting in an ensemble of $N_e$ pairs of control and perturbed trajectories. Importantly, for each pair of control and perturbed trajectories we set the same noise process by fixing the random seed.
    \item At each time lag $\tau$ we then estimate the time-dependent mean of an observable $\mathcal{A}(x_m(\tau))$ for both the perturbed and unperturbed runs. We refer to the perturbed and control ensemble averages as $\langle \mathcal{A}(x_m(\tau)) \rangle_{\text{p}}$ and $\langle \mathcal{A}(x_m(\tau)) \rangle$, respectively. The subscript $\text{p}$ in $\langle \rangle_{\text{p}}$ indicates ensemble averages over the perturbed distribution. The observables considered in this study are: (i) $\mathcal{A}(x_m(\tau)) = x_m(\tau)$ and (ii) $\mathcal{A}(x_m(\tau)) = (x_m(\tau) - \mu_m(\tau))^2$, where $\bm{\mu}(\tau)$ represents the time-dependent mean of the distribution. These observables quantify the responses of the ensemble mean and variance, respectively.
    \item We define the impulse response operator for observable $\mathcal{A}(x_m(\tau))$ as
    \begin{equation*}
        R^{m,l}_t = \frac{\langle \mathcal{A}(x_m(\tau)) \rangle_{\text{p}} - \langle \mathcal{A}(x_m(\tau)) \rangle}{\Delta_l}.
    \end{equation*}
\end{enumerate}

For the linear model we chose an ensemble size of $10^6$ members. Differently, the ensemble size for the numerical modified Lorenz--96 ground truth is $1.5 \times 10^8$: such large ensemble has been found to be needed to capture reliable responses in the ensemble variance.

\section{Linear approximation of the Lorenz--96 system at $J=1$} \label{app:experiment_J_1}

In the main text, the nearly Gaussian stationary marginals of the 20 resolved variables $x_k$ motivated our choice of a linear stochastic model for the case $J=500$. Here we fit the same linear model with memory (LMM) to the case $J=1$, with one fast variable per slow variable. In this case, the stationary marginals of $x_k$ are non-Gaussian, as is the joint distribution of $x_k$ and the coupling term $B_k$ shown in Figure~\ref{fig:Joint_PDFs}. We examine how accurately the LMM reproduces lagged correlations and ensemble-mean responses to forcing even when it misses the shape of the stationary marginals.

\subsection{Stationary statistics}

As in the main text, we fit the LMM in Eq.~\eqref{eq:LMM} using observations of the $K=20$ slow variables from a simulation of the modified Lorenz--96 system spanning $50{,}000$ model time units (MTU). We then simulate an LMM trajectory of the same length. We compare the stationary marginal distributions, the lagged autocorrelation of $x_k$, and the cross-correlations between $x_{k+d}(t+\tau)$ and $x_k(t)$ for $d=1,3,5,8$. The original system is translationally invariant, so we use $k=10$ as the reference site for the correlations.\\

\begin{figure}%[h!]
\centering
\includegraphics[width=1\linewidth]{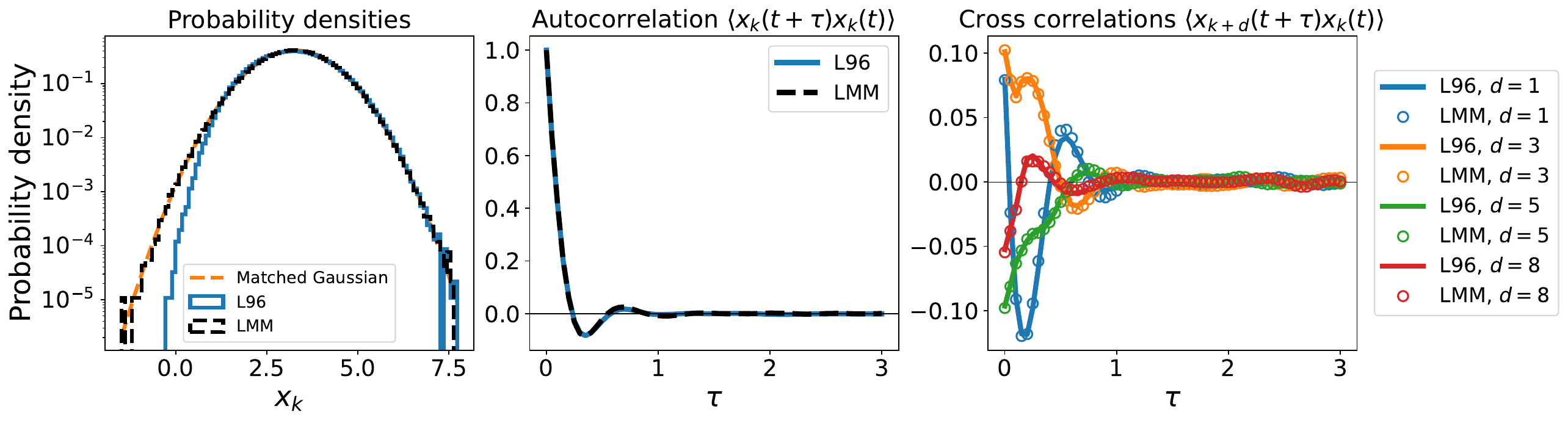}
\caption{Stationary statistics of the modified Lorenz--96 system (L96) and the linear model with memory (LMM) for $K=20$ and $J=1$. Left: marginal probability densities of $x_k$ for L96 and the LMM, compared with a Gaussian matching the L96 mean and variance. Center: lagged autocorrelation functions. Right: lagged cross-correlation functions between $x_{k+d}(t+\tau)$ and $x_k(t)$ for separations $d=1,3,5,8$. Lags $\tau$ are expressed in model time units.}
\label{fig:APP-PDF_correlations}
\end{figure}

Figure~\ref{fig:APP-PDF_correlations} shows that the LMM misses the non-Gaussian shape of the stationary marginal distribution, as expected. It nevertheless closely reproduces the lagged autocorrelation and the cross-correlations at all four spatial separations shown.

\subsection{Perturbed statistics}

\subsubsection{Linear response to an impulse perturbation}

We repeat the impulse experiment of Section~\ref{sec:response_operator} for $J=1$. At $t=0$, we perturb $x_{10}$ with a small impulse of size $0.1\sigma_x$, where $\sigma_x$ is the stationary standard deviation of the slow variables. We use \(10^7\) control--perturbed pairs for the modified Lorenz--96 system and \(10^6\) pairs for the LMM. Within each pair, both trajectories start from the same state before the impulse. We divide the resulting differences in ensemble mean and variance by the impulse amplitude. Details on the estimation of this response operator are in Appendix \ref{app:app_Response_Operator}.\\
Figure~\ref{fig:APP-Impulse_Response} shows that the LMM captures the main features of the ensemble-mean response even though the stationary marginals of the slow variables are non-Gaussian. The same impulse also produces a pronounced increase in ensemble variance around the perturbed site in the modified Lorenz--96 system. At \(k=10\) and \(\tau=0.20\) MTU, for example, the resulting change in standard deviation is comparable in magnitude to the change in the mean. In the LMM, linear dynamics and additive noise imply that the ensemble variance remains unchanged at every lag \cite{Majda2010}.

\begin{figure}
\centering
\includegraphics[width=\linewidth]{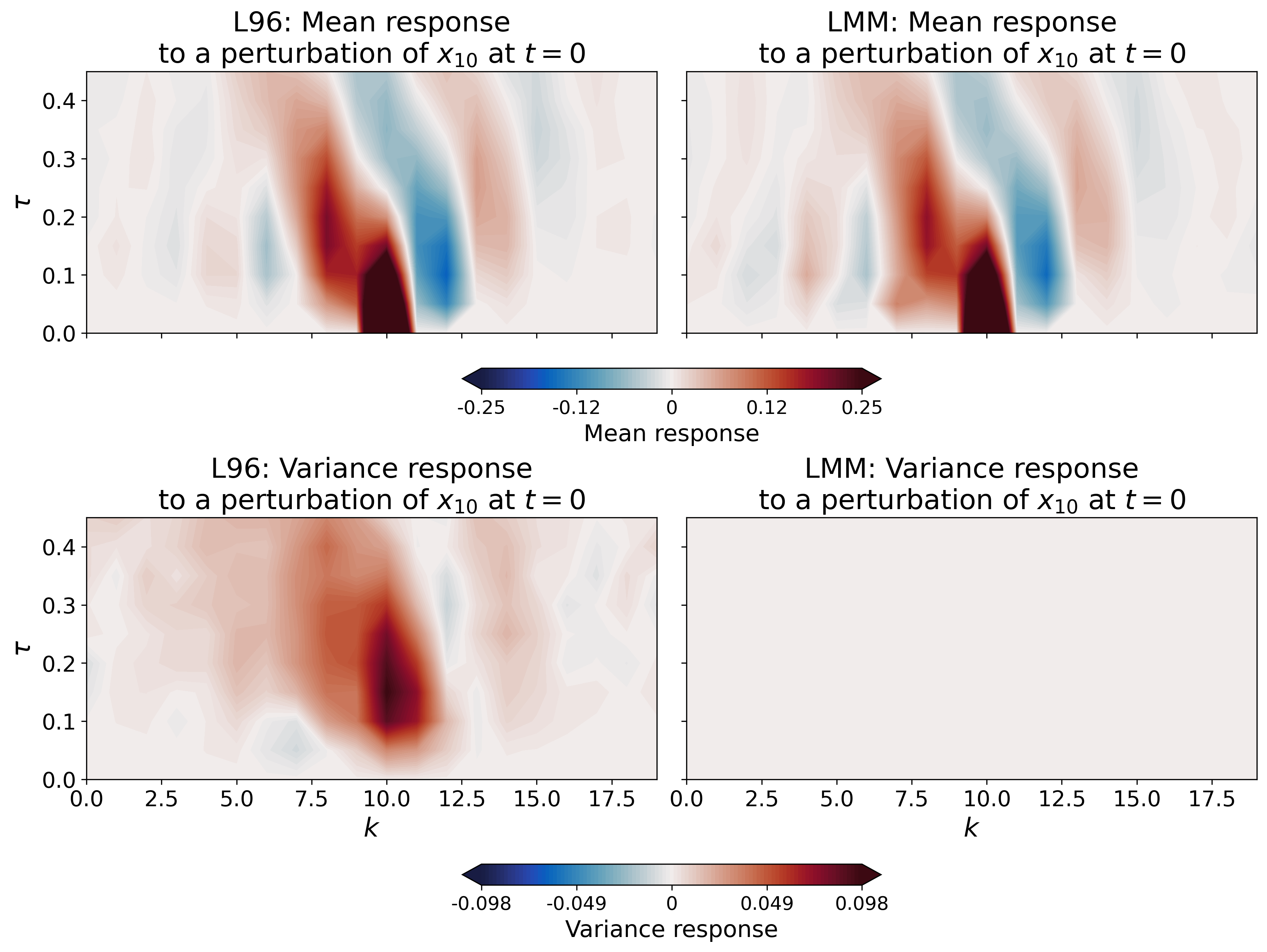}
\caption{Modified Lorenz-96 system: Linear response regime. Response operator $R_{m,l}(\tau)$ for the ensemble-mean response (top) and ensemble-variance response (bottom) to a small impulse applied to $x_{10}$ at $t=0$. The left column shows the modified Lorenz--96 system with $J=1$ and the right column shows the fitted LMM.}
\label{fig:APP-Impulse_Response}
\end{figure}

\subsubsection{Response to a large step forcing}

\begin{figure}
\centering
\includegraphics[width=\linewidth]{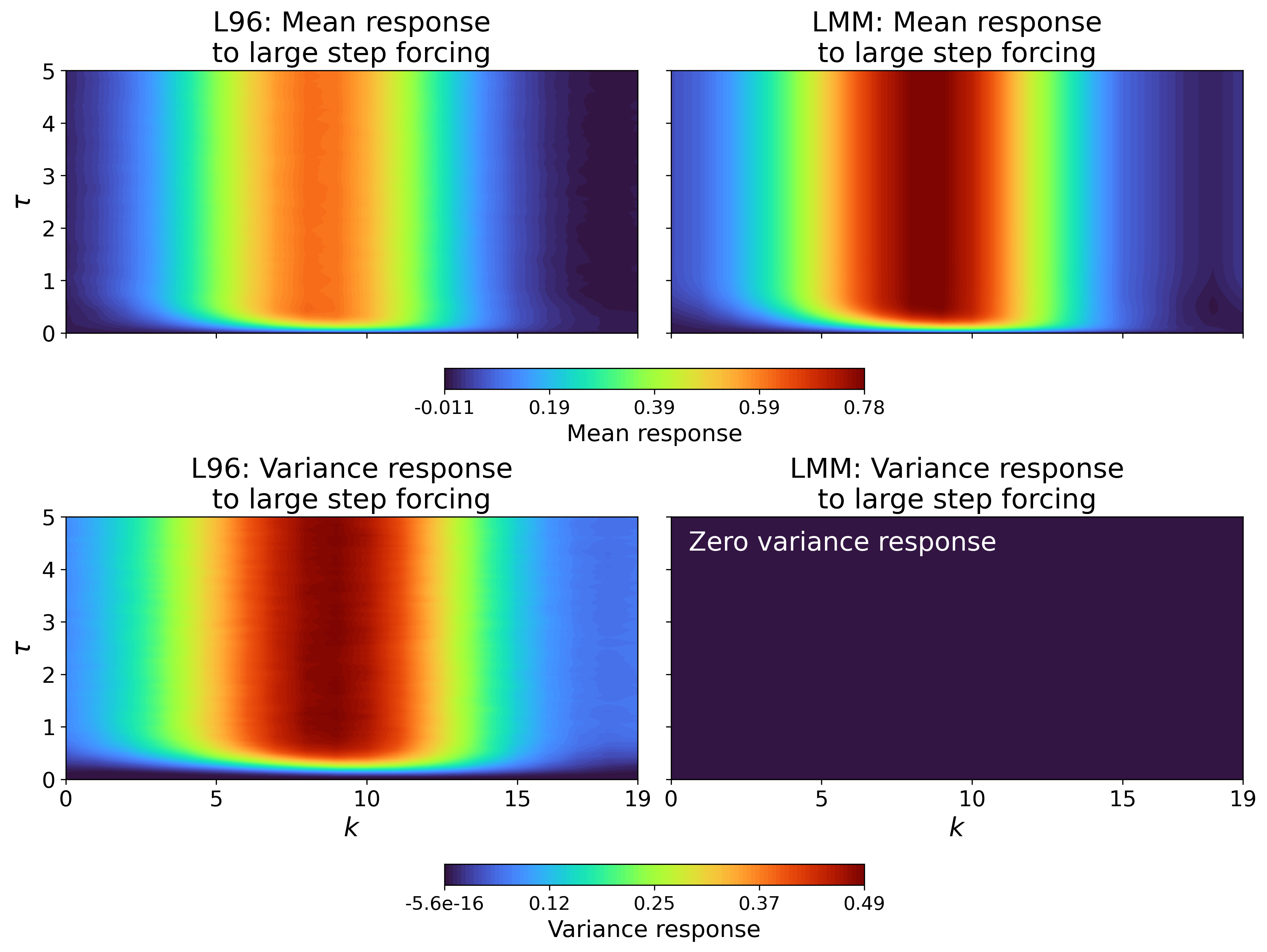}
\caption{Changes in ensemble mean (top) and ensemble variance (bottom) under the large, sustained Gaussian forcing in Eq.~\eqref{eq:gaussian_forcing}. The left column shows the modified Lorenz--96 system at $J=1$ and the right column shows the fitted LMM.}
\label{fig:APP-Step_Forcing}
\end{figure}

We apply the Gaussian-bump forcing defined in Eq.~\eqref{eq:gaussian_forcing} to the slow variables of both the modified Lorenz--96 system and the fitted LMM. As in the main text, the forcing is centered at $k_0=10$, has width $\ell=3$, and is held constant from $t=0$ onward. Its peak amplitude is $5\sigma_{\mathrm{ref}}$ per MTU, where $\sigma_{\mathrm{ref}}$ is the unforced standard deviation of $x_{10}$ at $J=1$. This is a severe forcing, with magnitude 5 times larger than the variability observed in the dataset. We use \(10^6\) ensemble members for each model and measure the changes in ensemble mean and variance relative to the corresponding unforced statistics.\\
Figure~\ref{fig:APP-Step_Forcing} shows that the LMM captures the main spatial pattern of the ensemble-mean response and its adjustment over time, although it overestimates the amplitude. The same forcing also changes the ensemble variance in the modified Lorenz--96 system. As in the impulse experiment, the LMM predicts zero variance response at every site and time.

\clearpage

\bibliographystyle{unsrtnat}
\bibliography{references}% Produces the bibliography via BibTeX.
\end{document}